\documentclass[manuscript,screen]{acmart}
\AtBeginDocument{%
  \providecommand\BibTeX{{%
    \normalfont B\kern-0.5em{\scshape i\kern-0.25em b}\kern-0.8em\TeX}}}

\setcopyright{acmcopyright}
\copyrightyear{2023}
\acmYear{2023}
\acmDOI{XXXXXXX.XXXXXXX}

\usepackage{graphicx}
\usepackage{subcaption}
\usepackage{multirow}
\usepackage{booktabs}
\usepackage{amsmath}
\usepackage{multicol}
\newtheorem{assumption}{Assumption}
\newtheorem{definition}{Definition}

\begin{document}

%\title{Integrating Relevance Judgement in Training Loss for Neural Sequential Recommenders}

\title{Integrating Item Relevance in Training Loss for Sequential Recommender Systems}

\author{Andrea Bacciu}
    \orcid{0009-0007-1322-343X}
    \affiliation{
      \institution{Sapienza University of Rome}
       \city{Rome}
      %\state{State}
      \country{Italy}
      }
    \email{bacciu@diag.uniroma1.it}

\author{Federico Siciliano}
    \orcid{0000-0003-1339-6983}
    \affiliation{
      \institution{Sapienza University of Rome}
       \city{Rome}
      %\state{State}
      \country{Italy}
      }
    \email{siciliano@diag.uniroma1.it}

\author{Nicola Tonellotto}
    \orcid{0000-0002-7427-1001}
    \affiliation{
      \institution{University of Pisa}
       \city{Pisa}
      %\state{State}
      \country{Italy}
      }
    \email{???}

\author{Fabrizio Silvestri}
    \orcid{0000-0001-7669-9055}
    \affiliation{
      \institution{Sapienza University of Rome}
       \city{Rome}
      %\state{State}
      \country{Italy}
      }
    \email{fsilvestri@diag.uniroma1.it}

% \affiliation{%


\begin{abstract}
\looseness -1 Sequential Recommender Systems (SRSs) are a popular type of recommender system that learns from a user's history to predict the next item they are likely to interact with.
However, user interactions can be affected by noise stemming from account sharing, inconsistent preferences, or accidental clicks.
To address this issue, we (i) propose a new evaluation protocol that takes multiple future items into account and (ii) introduce a novel relevance-aware loss function to train a SRS with multiple future items to make it more robust to noise.
Our relevance-aware models obtain an improvement of ~1.2\% of NDCG@10 and 0.88\% in the traditional evaluation protocol, while in the new evaluation protocol, the improvement is ~1.63\% of NDCG@10 and ~1.5\% of HR w.r.t the best performing models.

\end{abstract}




\maketitle

\section{Introduction}\label{sec:introduction}
\looseness -1 Recommender systems have become an integral part of our daily lives~\cite{zhang2019deep}, as they assist us in making decisions by suggesting items we might like based on our preferences and behaviors \cite{ricci2011introduction}. 
In recent years, Sequential Recommender Systems (SRSs) have emerged as a promising solution to improve the accuracy and relevance of recommendations by incorporating the temporal aspect of user-item interactions. 
These systems aim to predict the next item a user is likely to interact with based on their past interactions, taking into account the sequence of actions and their temporal order \cite{quadrana2018sequence}.

\looseness -1 SRSs have been successfully applied to various domains, including e-commerce \cite{
hwangbo2018recommendation}, music streaming \cite{schedl2015music}, and movie recommendations \cite{goyani2020review}.
Traditional evaluation metrics, such as Hit Rate (HR) and Normalized Discounted Cumulative Gain (NDCG), fail to capture the complexity of sequential data when evaluated by measuring how well they predict a single future item \cite{kang2018self,sun2019bert4rec,li2020time}. The reason is that the hypothesis of a single "relevant" item might not reflect the users' true intentions or preferences, particularly when considering noisy sequences in real-world scenarios \cite{wang2021denoising}. For example, users accidentally clicking on items, or performing multiple actions quickly, can greatly affect the evaluation results, as shown by~\citet{gupta2019dealing,oh2022rank}. The evaluation of SRSs has been thoroughly analyzed in recent research~\cite{sun2022counter,zhao2022revisiting,ji2023critical}. 
% \citet{oh2022rank} proposed new evaluation metrics and methods that consider multiple future items and incorporate the uncertainty and diversity of the recommendations.

Our paper contributes to this line of research by proposing a novel approach: we depart from the single-relevant item approach typically taken in the literature and adopt a novel evaluation method and training approach that considers multiple future items and accounts for noise in the sequences. The proposed method provides, as shown in the experiments, a more accurate and robust solution for evaluating and training SRSs.
\looseness -1 We conducted experiments on four datasets typically used in this research domain, using SASRec \cite{kang2018self}, a widely cited SRS, as the reference model for our experiments.

The research questions we address in our experiments are the following:
\begin{itemize}
    \item \textbf{RQ1:} Is there an alternative to the single relevant item evaluation protocol that is more suitable to determine the best-performing ranking model among several proposed options?
    \item \textbf{RQ2:} Can the item relevance be successfully incorporated into the training mechanism of an SRS to boost its performance?
    \item \textbf{RQ3:} What is the impact of the considered number of future items on the evaluation metrics as well as on the training performance of the models considered?
\end{itemize}

\looseness -1 Our paper makes two key contributions to the field of SRSs.
Firstly, we introduce a new evaluation protocol that considers multiple future items as potentially relevant instead of the typical single-relevant item hypothesis used in this research area.
Secondly, our experiments show that when trained in the multiple-relevant item regime, SASRec outperforms the state-of-the-art models, as shown by the significant improvements in NDCG@10 and Recall@10 scores.

\section{Related Work}

\subsection{Sequential Recommender Systems}
Sequential Recommender Systems (SRSs) are a class of recommender systems that personalizes recommendations to users based on their historical interactions with items in a sequence \cite{wang2019sequential}, capturing its temporal dynamics.
SRSs have received considerable attention in the research community in recent years \cite{zhang2019deep} and have been applied in various domains \cite{zhang2019deep}, including movies \cite{harper2015movielens, goyani2020review, petrov2022effective}, music \cite{schedl2015music,schedl2018current}, and e-commerce \cite{schafer2001commerce,hwangbo2018recommendation}.
Various techniques have been developed to model the temporal dependencies in the sequences, including Markov Chain models, Recurrent Neural Networks (RNNs), and Attention mechanisms.
Markov Chain models are a type of probabilistic model that assume the future state of a sequence only depends on the current state; for this reason, they struggle to capture complex dependencies in long-term sequences \cite{fouss2005novel,fouss2005web}.
RNNs are a type of neural network architecture that can capture the long-term dependencies in sequential data.
They have shown great potential in modeling sequential data, and they have been used to develop various SRSs, such as session-based recommenders \cite{hidasi2015session,hidasi2016parallel,li2017neural,hidasi2018recurrent,liu2020adversarial}, context-aware recommenders \cite{adomavicius2010context,yang2018lstm,kulkarni2020context}, and sequential graph neural networks \cite{chang2021sequential,fan2021continuous}.
Attention mechanisms have recently gained attention in SRSs due to their ability to dynamically weight the importance of different parts of the sequence \cite{kang2018self}. By doing so, attention mechanisms can better capture the important features in the sequence and improve the prediction accuracy \cite{zhou2018atrank, wang2019neural}.

\subsection{Evaluating Sequential Recommender Systems}
Evaluating Sequential Recommender Systems (SRSs) has been a topic of great interest in recent years.
Both \cite{sun2022counter} and \cite{zhao2022revisiting} examined common data splitting methods for SRSs and discussed why commonly used evaluation methods are ill-defined, suggesting appropriate offline evaluation for SRSs. In particolar, they showed that existing evaluation protocols do not consider the temporal dynamics of user behaviour, which can affect the accuracy of the recommendations.
In \cite{ji2023critical} it is shown that the current evaluation protocols for SRSs can lead to data leakage, where the model learns information from the test data that is not available during training. They address the problem by proposing an evaluation methodology that takes into account the global timeline of data samples in the evaluation of SRSs.
A metric called Rank List Sensivity (RLS) is introduced in \cite{oh2022rank} to evaluate the discrepancy between two rankings, so to evaluate models' sensitivity with respect to training data.
Finally, \cite{baraglia2009search} presented an evaluation methodology specifically designed to evaluate the precision of algorithms for the Search Shortcut Problem. This metric considers item relevance, which allows an effective evaluation.

\section{Methodology}
\subsection{Current Evaluation Protocol}
In line with previous research employing SRSs~\cite{kang2018self,sun2019bert4rec,li2020time}, the current evaluation method involves shuffling one positive item (the next item in the sequence) with 100 random negative items not part of the input sequence. These items are then ranked based on their relevance scores determined by the model. The resulting rank is used to calculate evaluation metrics like Normalized Discounted Cumulative Gain (NDCG) and Hit Rate (HR), which can be measured with various cut-offs, typically 10.

\subsection{Problems with the current protocol}\label{sec:problem-current-protocol}
The current evaluation protocol assumes only one item is relevant to a user, which may not be true in real-world scenarios where multiple interactions could be relevant. Users' history can be noisy, as shown in \cite{wang2021denoising}. Noise can come from account sharing, inconsistent preferences, or accidental clicks. For instance, in e-commerce, many clicks don't lead to purchases, and some purchases receive negative reviews. Evaluating a model based on one positive item, which might be an error, negatively impacts its performance, and this issue is not addressed in the current evaluation protocol.

\subsection{Multi Future Items: a new evaluation protocol}
To address the aforementioned problem, we propose a new evaluation protocol for SRSs called Multi Future Items (MFI). In MFI, we make the Assumption \ref{asm:eval_protocol}, that a good ranking should not only contain the single future next item in the sequence, but the whole sequence of future items in the correct order. Hence, MFI is rewarding the models that produce a better ranking considering multiple future items.

\begin{assumption}\label{asm:eval_protocol}
Given a user $u$ and its ordered interactions' sequence $[I_1, I_2, ..., I_i]$, the ideal ranking of length $K$ is the sequence of future items $[I_{i+1}, I_{i+2},..., I_{i+K}]$ arranged user's interaction temporal order.
\end{assumption}

To evaluate the performance of SRSs under Assumption \ref{asm:eval_protocol}, we use traditional evaluation metrics for sequential recommendation models such as NDCG and HR. To have multiple future items per evaluation, we therefore split the user's history differently. Given a sequence $[I_1, I_2, ..., I_L]$, in the traditional evaluation protocol only item $I_L$ is reserved for testing. In our proposed evaluation protocol, the sequence $[I_1, I_2, ..., I_{L-K}]$ is allocated for training the model, while the sequence $[I_{L-K+1}, I_{L-K+2}, ..., I_L]$ is used for testing purposes.

%La nostra idea di usare le item relevance consiste nel cambiare il training, ma non l'evaluation. perché non avrebbe senso

\subsection{Item Relevance}\label{sec:item-relevance}
\begin{figure}
    \centering
    \includegraphics[width=0.7\textwidth]{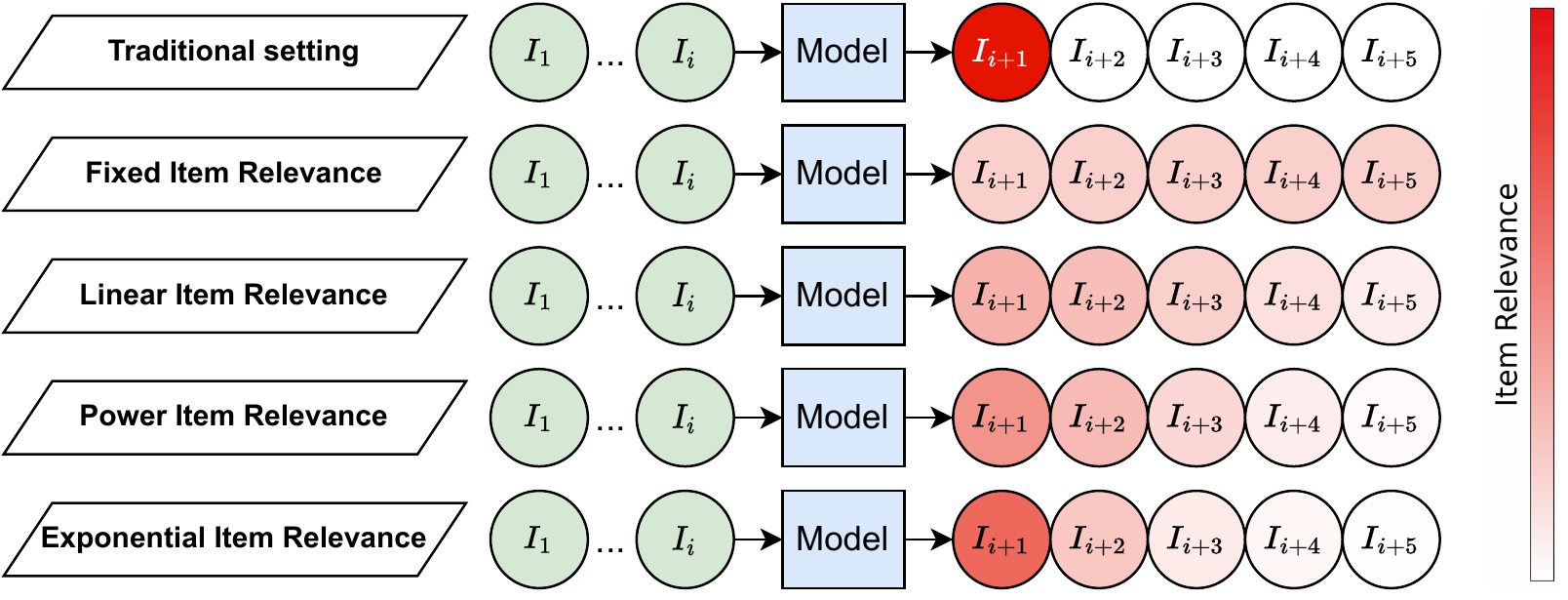}
    \caption{A visualization of how the various loss scaling strategy weigh the item relevance.}
    \label{fig:item_relevance}
\end{figure}

\looseness -1 The new evaluation protocol requires higher ranking capabilities than the traditional evaluation protocol.
In the original evaluation protocol, the ability to rank a single item and treat all other items as not relevant to the user is evaluated.
Conversely, in the new evaluation protocol, it is important to define item relevance to give importance to multiple future items and scale their importance according to their position in the sequence.

\begin{definition}\label{def:item_rel_function}
Given a ranking $[I_{1}, I_{2},..., I_{K}]$ of length $K$, we define the item relevance function $r \colon$
$\mathbb{N} \rightarrow [0,1]$
\end{definition}

\looseness -1 In order to compare the relevance of items on sequences of different lengths $K$, we further define that the item relevance sums to one.

\begin{definition}\label{def:item_rel_sum}
Given a ranking $[I_{1}, I_{2},..., I_{K}]$ of length $K$, $r$ must satisfy $\sum_{i=1}^K r(i) = 1$
\end{definition}

\looseness -1 Item relevance should account for the fact that some interactions will occur further in the future and therefore assign higher importance to them: we establish that the relevance of an item at time $t$ cannot be lower than the relevance of the item at time $t+1$.

\begin{definition}\label{def:item_rel_greater}
Given a ranking $[I_{1}, I_{2},..., I_{K}]$ of length $K$, $r$ must satisfy $r(i+1) \leq r(i) \quad \forall i \in \{1,2,...,K\}$
\end{definition}

Our approach to item relevance is inspired by \cite{baraglia2009search} who proposed a similarity function, which takes into account the item relevance, to evaluate query recommendation using collaborative filtering. They suggested four different functions for item relevance: $r(i) = 1$, $r(i) = K-i$, $r(i) = \left(K-i\right)^2$, and $r(i) = e^{K-i}$ with $i \in \{1,2,...,K\}$. 
We refer to these functions respectively as \textit{Fixed}, \textit{Linear}, \textit{Power}, and \textit{Exponential}.
The first assigns equal relevance to all items, while the others assign higher relevance to the next items in the sequence at the expense of the more distant ones. We show a visualization of these functions in Figure \ref{fig:item_relevance}.
The functions can be easily normalized to comply with the Definition \ref{def:item_rel_sum}.
It should be noted that our evaluation protocol generalizes the traditional one because we can revert to it by setting a maximum importance to the next item and zero importance for all other future items.

\subsection{Relevance-based Loss}
\looseness -1 Current Neural SRSs use Binary Cross-Entropy as the loss function; we propose a modification of it called Relevance-based loss.
Relevance-based loss uses multiple positive items, similarly to what proposed by \cite{tang2018personalized}, except that it assigns weights to them following the functions defined in Section \ref{sec:item-relevance}.
%\cite{}, on the other hand, rather than weighing the data, directly samples it based on recency.
\cite{tang2018personalized}, though, propose a formulation that assigns equal relevance to all future items (equivalent to our Fixed formulation). However, by assigning equal importance to all items, the model ignores the natural order of interactions, making this strategy unsuitable for some tasks.
For instance, in the case of movie recommendations, it is not realistic to suggest \textit{Back To The Future 3} before the user has watched the first and the second.
To address this limitation, we propose a more general formula that accounts for item relevance through time, which we call Relevance-based loss:

\begin{equation} \label{eq:cross_entropy_pos}
    \ell(\vec x \mid pos, neg, r) = -\sum_{i = 1}^{pos} \text{log} \left(\left(\vec x_{pos}\right)_i\right)r(pos-i+1) - \sum_{i = 1}^{neg} \text{log} \left(1-\left(\vec x_{neg}\right)_i\right)
\end{equation}

where $pos$ and $neg$ represent respectively the number of positive and negative items considered. $\vec x$ is the score given by the model to each item, while $\vec x_{pos}$ and $\vec x_{neg}$ represent respectively the subset of $\vec x$ containing only the positive and negative items and $r$ is the item relevance function defined in Section \ref{sec:item-relevance}.
In Equation \ref{eq:cross_entropy_pos}, we weigh the loss of each item $i$ by its relevance score $r(i)$. By doing so, the model is encouraged to focus more on the relevant items while learning.
%The relevance score is determined by our earlier assumption that the relevance of a future item cannot be lower than the relevance of the item that comes after it.

%\looseness -1 To demonstrate the effectiveness of this approach, we evaluate the proposed relevance judgement loss with different relevance functions: $r(x)=1$, $r(x)=x$, $r(x)=x^2$ and $r(x)=e^x$. Using $r(x) = 1$ gives us the same equation as \cite{tang2018personalized}'s loss)

\section{Experimental Setup}\label{sec:dataset}
We assess our techniques using four datasets derived from real-world use cases: MovieLens\cite{harpermovielens} and Foursquare \cite{yang2014modeling}. MovieLens is a popular benchmark dataset for Recommendation Systems containing movie ratings. We utilize both MovieLens-1M and MovieLens-100K, which comprises 1 million and 100.000 user ratings, respectively. The Foursquare NYC and Foursquare TKY datasets, presented in \cite{yang2014modeling}, are collections of check-in records from New York and Tokyo. The NYC dataset contains 227,428 records, while the TKY dataset has 573,703 records.

\looseness -1 We select the Self-Attentive Sequential Recommendation (SASRec) model\footnote{\url{https://github.com/pmixer/SASRec.pytorch}} \cite{kang2018self} for our experiment, as it has consistently demonstrated exceptional performance across multiple benchmarks and garnered significant recognition in the literature.
To have a fair comparison, we keep the original hyper-parameter of SASRec's paper.
We perform our experiments on a workstation equipped with an Intel Core i9-10940X (14-core CPU running at 3.3GHz) and 256GB of RAM, and a single Nvidia RTX A6000 with 48GB of VRAM.

\section{Results}\label{sec:results}
In this Section, we present the results to answer our posed
research questions (Section \ref{sec:introduction}).
During our experiments, we vary (i) The number of evaluation positives; (ii) The number of training positives; (iii) the item relevance function.
From here on, we will denote by 'our' o 'new' models, the models trained with our proposed training loss.

\subsubsection*{\textbf{RQ1:} Is there an alternative to the single relevant item evaluation protocol that is more suitable to determine the best-performing ranking model among several proposed options?\\}
Table \ref{tab:complete_results_1} compares various models in the traditional evaluation protocol, including two baselines (Baseline and Fixed) and the models trained with our proposed item relevance-based loss (Linear, Power, Exponential). 
As expected, in the traditional evaluation protocol, there is no clear winner.
Instead, in Table \ref{tab:complete_results_10}, the performance discrepancies are more pronounced and it is, therefore, possible to identify a better model.
Hence, the new evaluation protocol is more robust to the noise introduced in \ref{sec:problem-current-protocol} and can better identify performance differences between models and reward the model with the best ranking performance as evaluated by means of NDCG and HR.
\looseness -1 We can also assert that models that discount the future excessively, i.e. only (or almost only) take into account the next item (Baseline and Exponential), generally perform worse.

\begin{table}[hbtp]
    \centering
    \begin{tabular}{lcccccccccc}
        \toprule
        & \multicolumn{2}{c}{ML-1M} & \multicolumn{2}{c}{ML-100k} & \multicolumn{2}{c}{Foursquare NYC} & \multicolumn{2}{c}{Foursquare TKY} \\
        \midrule
        Model & NDCG & HR & NDCG & HR & NDCG & HR & NDCG & HR \\
        \midrule\midrule
        Baseline~\cite{kang2018self} & \underline{0.5855} & \textbf{0.8228} & 0.4300 & 0.7041 & 0.6649 & 0.7525 & 0.7217 & \textbf{0.8116} \\
        Fixed~\cite{tang2018personalized} & 0.5483 & 0.8017 & 0.4246 & \underline{0.7084} & 0.6698 & \underline{0.7581} & \underline{0.7247} & \underline{0.8094} \\
        \midrule
        Linear & \textbf{0.5896} & 0.8187 & \textbf{0.4358} & \textbf{0.7137} & \underline{0.6715} & \underline{0.7581} & \textbf{0.7295} & 0.8081 & \\
        Power & 0.5751 & \underline{0.8220} & \underline{0.4309} & 0.6999 & \textbf{0.6765} & \textbf{0.7719} & 0.7229 & 0.8090 \\
        Exponential & 0.5169 & 0.7763 & 0.4233 & 0.6999 & 0.5717 & 0.7442 & 0.6195 & 0.7872 \\
        \bottomrule
    \end{tabular}
    \caption{The table shows the values of the NDCG@10 (higher is better) and HR@10 (higher is better) metrics that the five models obtain on the four datasets in the traditional evaluation protocol. \textbf{Bold} shows the best result for each column, \underline{underlined} the second best.}
    \label{tab:complete_results_1}
\end{table}

\begin{table}[htbp]
    \centering
    \begin{tabular}{lcccccccccc}
        \toprule
        & \multicolumn{2}{c}{ML-1M} & \multicolumn{2}{c}{ML-100k} & \multicolumn{2}{c}{Foursquare NYC} & \multicolumn{2}{c}{Foursquare TKY} \\
        \midrule
        Model & NDCG & HR & NDCG & HR & NDCG & HR & NDCG & HR \\
        \midrule\midrule
        Baseline~\cite{kang2018self} & 0.6532 & 0.6151 & 0.5638 & 0.5269 & 0.7617 & \textbf{0.6864} & 0.8080 & 0.7433 \\
        Fixed~\cite{tang2018personalized} & 0.6562 & 0.6188 & 0.5632 & 0.5266 & 0.7611 & \underline{0.6848} & 0.8139 & 0.7508 \\
        \midrule
        Linear & \textbf{0.6694} & \textbf{0.6307} & \textbf{0.5753} & \textbf{0.5368} & 0.7549 & 0.6765 & \textbf{0.8234} & \textbf{0.7570} \\
        Power & \underline{0.6607} & \underline{0.6230} & \underline{0.5694} & \underline{0.5298} & \textbf{0.7625} & \underline{0.6848} & \underline{0.8193} & \underline{0.7549} \\
        Exponential & 0.6536 & 0.6147 & 0.5667 & 0.5297 & \underline{0.7624} & 0.6845 & 0.8185 & 0.7544 \\
        \bottomrule
    \end{tabular}
    \caption{The table shows the values of the NDCG@10 (higher is better) and HR@10 (higher is better) metrics that the five models obtain on the four datasets in the new evaluation protocol using 10 positive items. \textbf{Bold} shows the best result for each column, \underline{underlined} the second best.}
    \label{tab:complete_results_10}
\end{table}

\subsubsection*{\textbf{RQ2:} Can the item relevance be successfully incorporated into the training mechanism of an SRS to boost its performance?\\}
As can be seen from Table \ref{tab:complete_results_1}, in the traditional evaluation protocol, models that integrate the item relevance are on-par or better than the baselines for at least one of the metrics. In particular, Linear item relevance outperforms the other approaches in 6 out of 10 cases while getting the second-best result in 2 cases.
When considering more positive items for evaluation (see Table \ref{tab:complete_results_10}) it is possible to notice that the Linear model outperforms the baselines and the other models, except in a benchmark where Power item relevance yields a better NDCG score and the baselines a better HR. Fixed, on the other hand, returns performance comparable to Baseline, suggesting that incorporating more positive items during training, even if without weighting them in the loss, can lead to a slight increase in performance.

\begin{figure}
     \centering
     \begin{subfigure}[t]{0.49\textwidth}
         \centering
         \includegraphics[width=\textwidth]{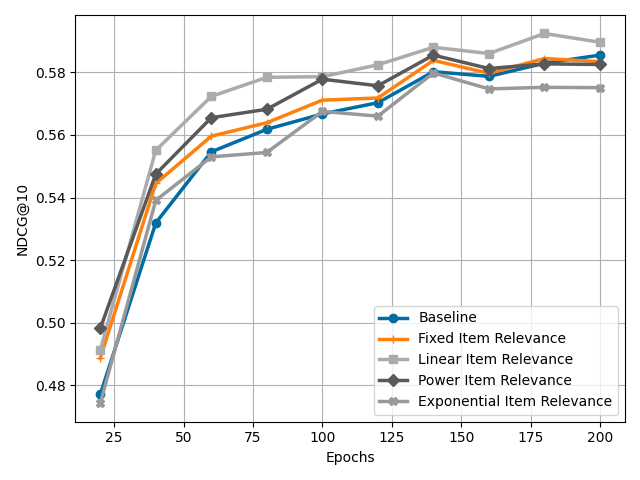}
         \caption{Traditional evaluation protocol}
         \label{fig:epoch_ml-1m_1}
     \end{subfigure}
     \hfill
     \begin{subfigure}[t]{0.49\textwidth}
         \centering
         \includegraphics[width=\textwidth]{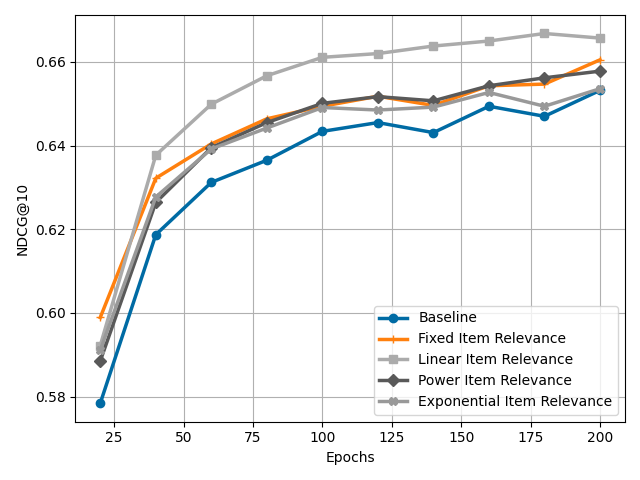}
         \caption{New evaluation protocol}
         \label{fig:epoch_ml-1m_10}
     \end{subfigure}
    \caption{NDCG@10 at varying training epochs for all 5 models on the MovieLens 1M dataset}
    \label{fig:number_removal_comparison}
\end{figure}

Figure \ref{fig:number_removal_comparison} shows the performance using NDCG@10 for the MovieLens-1M dataset for both the traditional and new evaluation protocol. From Figure \ref{fig:number_removal_comparison}, the improvement provided by item-relevance loss is more evident. We can in fact see that the Baseline (the original SASRec), which does not consider multiple future items during training, has a consistently slower convergence rate in both evaluation protocols than the item relevance-aware models. Instead, Linear item relevance obtains the fastest convergence.
This shows that the item relevance loss allows the training of better models that yield higher performances in both evaluation protocols.

To summarize, our relevance-aware models obtain an improvement of ~1.2\% of NDCG@10 and 0.88\% in the traditional evaluation protocol, while in the new evaluation protocol, the improvement is ~1.63\% of NDCG@10 and ~1.5\% of HR.
%Similar results are observed using the other datasets; however, for the sake of brevity, we report these results in the Github repository.

\subsubsection*{\textbf{RQ3:} What is the impact of the considered number of future items on the evaluation metrics as well as on the training performance of the models considered?\\}
\begin{figure}[htbp]
    \centering
    \includegraphics[width=0.55\textwidth]{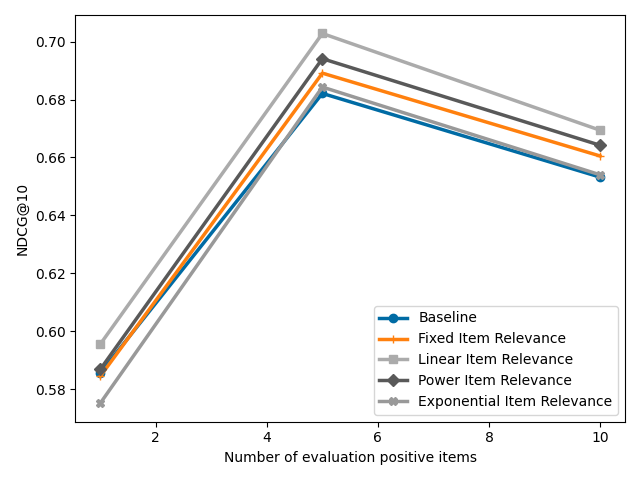}
    \caption{NDCG@10 at varying number of positive items used for evaluation on Movielens-1M}
    \label{fig:eval_positives_comparison}
\end{figure}

\looseness -1 To assess this question, we conduct two ablation studies, one in which we vary the number of evaluation positive items and another in which we vary the number of training positive items.
In Figure \ref{fig:eval_positives_comparison}, we report the results of all the models using 1, 5, and 10 evaluation positive items. We can notice that the relative performance of the relevance-aware models remains consistent when varying numbers of evaluation positive items, despite variations in their absolute performance.

\begin{figure}
     \centering
     \begin{subfigure}[t]{0.49\textwidth}
         \centering
         \includegraphics[width=\textwidth]{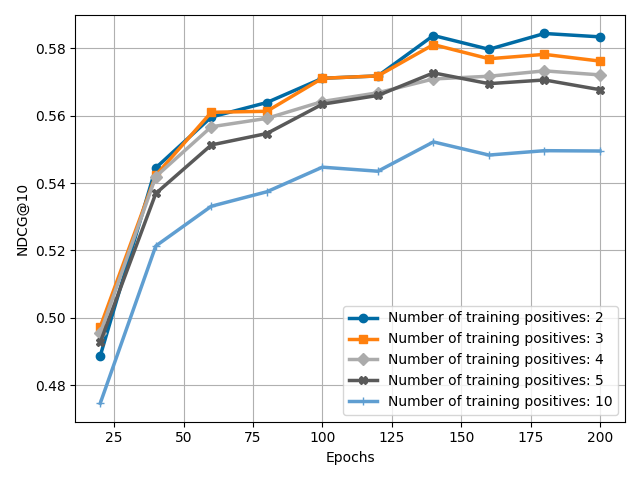}
         \caption{Traditional evaluation protocol - Fixed Item Relevance}
         \label{fig:train_positives_fixed_1}
     \end{subfigure}
     \hfill
     \begin{subfigure}[t]{0.49\textwidth}
         \centering
         \includegraphics[width=\textwidth]{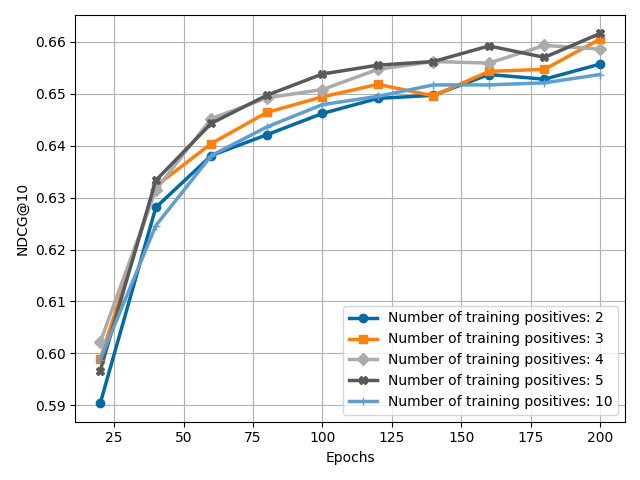}
         \caption{New evaluation protocol - Fixed Item Relevance}
         \label{fig:train_positives_fixed_10}
     \end{subfigure}
     \\
     \begin{subfigure}[t]{0.49\textwidth}
         \centering
         \includegraphics[width=\textwidth]{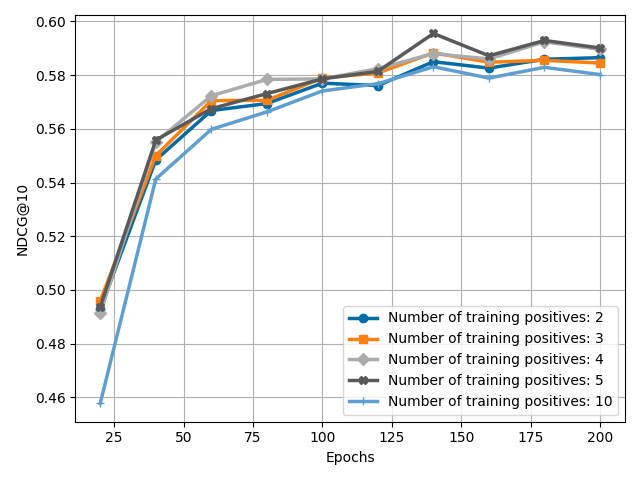}
         \caption{Traditional evaluation protocol - Linear Item Relevance}
         \label{fig:train_positives_linear_1}
     \end{subfigure}
     \hfill
     \begin{subfigure}[t]{0.49\textwidth}
         \centering
         \includegraphics[width=\textwidth]{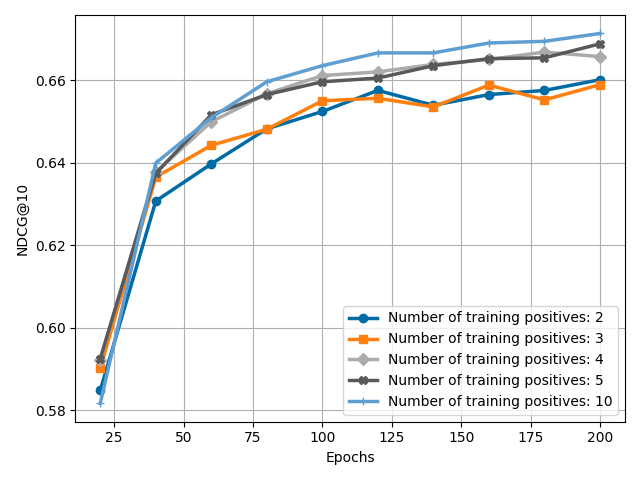}
         \caption{New evaluation protocol - Linear Item Relevance}
         \label{fig:train_positives_linear_10}
     \end{subfigure}
    \caption{NDCG@10 at varying training epochs for the new item-relevance models for different numbers of training positives on Movielens-1M}
    \label{fig:train_positives_comparison}
\end{figure}

In Figure \ref{fig:train_positives_comparison} we report the results of all models using 2, 3, 4, 5, 10 training positive items.
In the traditional evaluation protocol (Figures \ref{fig:train_positives_fixed_1}, \ref{fig:train_positives_linear_1}), increasing the number of training positive items seems to impact the performances negatively: increasing the number of training positive items leads to a deterioration of performance. This is particularly evident for Fixed Item Relevance, where an increase in training positive items always leads to a decrease in performance. For Linear, this behaviour is more attenuated, although using 10 training positive items still generates the worst performance.
In the traditional evaluation protocol, the behaviour is generally expected, as we are only evaluating the model with one positive item; the other positive items used for training can only confound the model.
Instead, in the new evaluation protocol (Figures \ref{fig:train_positives_fixed_10}, \ref{fig:train_positives_linear_10}), where we are evaluating models with 10 positive items, we see something different. Fixed Item Relevance increases performance up to 5 training positives, but at 10 it performs worst. This indicates that this type of relevance function is not particularly suitable, even when the training is similar to the evaluation.
Conversely, Linear shows interesting results: performance increases with the number of training positives. This suggests that our models still have room for further improvement.
Although not shown, the results for Power Item Relevance are similar to those for Linear, while Exponential shows a similar trend to Fixed in general.

\section{Conclusions}
%In this work, we show how much is important the user item relevance and how they boost 
In this work, we challenged the typical assumption made in Sequential Recommendation Systems (SRSs), which only consider the immediate next item in a sequence as the one to predict.
We have relaxed the evaluation protocol to not penalize the model in case of noisy sequences and designed an item relevance loss in order to optimize the model to predict multiple future items.
We demonstrated the importance of having more positive examples both in the training and evaluation of SRSs.
Our experiments show that when trained in the multiple relevant item regime, our systems outperform the state-of-the-art models, with an improvement of ~1.2\% of NDCG@10 and 0.88\% in the original evaluation protocol. In the new evaluation protocol, the improvement is ~1.63\% of NDCG@10 and ~1.5\% of HR. Among all the variants of grading multiple relevant items that we experimented with, it turns out that the Linear one outperforms the others and demonstrates its potential usefulness in practical applications.

\begin{acks}
This work was partially supported by projects FAIR (PE0000013) and SERICS (PE00000014) under the MUR National Recovery and Resilience Plan funded by the European Union - NextGenerationEU and by ERC Starting Grant No. 802554 (SPECGEO) and PRIN 2020 project n.2020TA3K9N "LEGO.AI".
\end{acks}

\bibliographystyle{ACM-Reference-Format}
\bibliography{citations}

\end{document}